\RequirePackage{lineno}
\documentclass[aps,twocolumn,showpacs,byrevtex,prl,reprint]{revtex4-1}

\usepackage{graphicx}
\usepackage{dcolumn}
\usepackage{bm}
\usepackage{rotating}
\usepackage{epstopdf}
\usepackage{color}
\usepackage{verbatim} 
\usepackage{multirow}
\usepackage[abs]{overpic}
\usepackage{amsmath}
\usepackage{amssymb}
\usepackage{xspace}

\newcommand{\PreserveBackslash}[1]{\let\temp=\\#1\let\\=\temp}
\newcolumntype{C}[1]{>{\PreserveBackslash\centering}p{#1}}
\newcolumntype{R}[1]{>{\PreserveBackslash\raggedleft}p{#1}}
\newcolumntype{L}[1]{>{\PreserveBackslash\raggedright}p{#1}}

\newcommand{\EE}{e^+e^-}

\newcommand{\too}{\rightarrow}

\begin{document}
\graphicspath{{figure/}}
\DeclareGraphicsExtensions{.eps,.png,.ps}
\title{\boldmath Cross section measurements of $e^+e^- \rightarrow \omega\chi_{c0}$ from $\sqrt{s} = 4.178$ to $4.278$ GeV}
\author{
  \begin{small}
    \begin{center}
      M.~Ablikim$^{1}$, M.~N.~Achasov$^{10,d}$, P.~Adlarson$^{59}$, S. ~Ahmed$^{15}$, M.~Albrecht$^{4}$, M.~Alekseev$^{58A,58C}$, A.~Amoroso$^{58A,58C}$, F.~F.~An$^{1}$, Q.~An$^{55,43}$, Y.~Bai$^{42}$
% ...
      \\
      \vspace{0.2cm}
      (BESIII Collaboration)\\
      \vspace{0.2cm} {\it
        $^{1}$ Institute of High Energy Physics, Beijing 100049, People's Republic of China\\
% ...
$^{4}$ Bochum Ruhr-University, D-44780 Bochum, Germany\\
% ...
$^{10}$ G.I. Budker Institute of Nuclear Physics SB RAS (BINP), Novosibirsk 630090, Russia\\
% ...
$^{15}$ Helmholtz Institute Mainz, Johann-Joachim-Becher-Weg 45, D-55099 Mainz, Germany\\
% ...
$^{42}$ Southeast University, Nanjing 211100, People's Republic of China\\
$^{43}$ State Key Laboratory of Particle Detection and Electronics, Beijing 100049, Hefei 230026, People's Republic of China\\
% ...
$^{55}$ University of Science and Technology of China, Hefei 230026, People's Republic of China\\
% ...
$^{58}$ (A)University of Turin, I-10125, Turin, Italy; (B)University of Eastern Piedmont, I-15121, Alessandria, Italy; (C)INFN, I-10125, Turin, Italy\\
$^{59}$ Uppsala University, Box 516, SE-75120 Uppsala, Sweden\\
% ...
$^{d}$ Also at the Novosibirsk State University, Novosibirsk, 630090, Russia\\
% ...
      }\end{center}
    \vspace{0.4cm}
\end{small}
}
\affiliation{}


\begin{abstract}
The cross section of the process $e^+e^- \rightarrow \omega \chi_{c0}$ is measured at center-of-mass energies from $\sqrt{s} =$ 4.178 to 4.278 GeV using a data sample of 7 fb$^{-1}$ collected with the BESIII detector operating at the BEPCII storage ring. The dependence of the cross section on $\sqrt{s}$ shows a resonant structure with mass of $(4218.5\pm1.6(\text{stat.})\pm4.0(\text{syst.}))$~MeV/$c^2$ and width of $(28.2\pm3.9(\text{stat.})\pm1.6(\text{syst.}))$~MeV, respectively. This observation confirms and improves upon the result of a previous study. The angular distribution of the $e^+e^- \rightarrow \omega \chi_{c0}$ process is extracted for the first time.
\end{abstract}


\maketitle
\section{I. INTRODUCTION}
$Y(4260)$ is the first charmonium-like $Y$ state, which was observed in the process $\EE \too \pi^+\pi^-J/\psi$ by the BABAR experiment using an initial-state-radiation (ISR) technique~\cite{Y4260-babar}. This observation was immediately confirmed by the CLEO~\cite{Y4260-cleo} and Belle experiments~\cite{Y4260-belle} in the same process. $Y(4360)$ and $Y(4660)$ were also observed in $\EE \too \pi^+\pi^-\psi(3686)$~\cite{Y4360-babar, Y4360-belle}. The observation of these $Y$ states has stimulated substantial theoretical discussions on their nature~\cite{Y-theory1, Y-theory2}. These $Y$ states do not fit in the conventional charmonium spectroscopy, so they are good candidates for exotic states, such as hybrid states, tetraquark states and molecule states~\cite{Y-theory3}. BESIII recently investigated the process $\EE \too \omega\chi_{c0}$ using data collected at $\sqrt{s}$ = 4.23 and 4.26~GeV combined with smaller data samples at nearby energies~\cite{omegachic}.  An enhancement was found in the cross section around $\sqrt{s}$ = 4.22 GeV, referred to as the $Y(4220)$. Resonance signals were not observed in a subsequent study above 4.4 GeV~\cite{omegachic2}. Various models~\cite{the1,the2,the3,the4,the5,the6,the7} are proposed to explain the observed line-shape. Possible scenarios include a missing $\psi(4S)$ state~\cite{the3}, a contribution from the $\psi(4160)$ state~\cite{the4}, a tetraquark state~\cite{the5}, or a molecule state~\cite{the6,the7}. Intriguingly, similar and possibly related structures are also observed in the same energy region for other processes, such as $\EE \too \pi^+\pi^-J/\psi$~\cite{pipijpsi-bes}, $\EE \too \pi^+\pi^-\psi(3686)$~\cite{pipipsip-bes}, $\EE \too \pi^+\pi^-h_c$~\cite{pipihc-bes}, and $\EE \too \pi^{+}D^{0}D^{*-}+c.c.$~\cite{piDDstar}.

In this paper, we report a study of the $\EE\too\omega\chi_{c0}$ reaction based on the most recent $\EE$ annihilation data collected with the BESIII detector~\cite{besiii} at nine energy points in the range 4.178 $\leqslant\sqrt{s}\leqslant$ 4.278 GeV, with a total integrated luminosity of about 7 fb$^{-1}$. The $\chi_{c0}$ state is detected via $\chi_{c0} \too \pi^+\pi^-/K^+K^-$, and the $\omega$ is reconstructed via the  $\omega \too \pi^{+}\pi^{-}\pi^{0}$ decay.

\section{II. BESIII DETECTOR AND MONTE CARLO SIMULATION}
The BESIII detector is a magnetic spectrometer~\cite{besiii} located at the Beijing Electron
Positron Collider (BEPCII)~\cite{bepcii}. The
cylindrical core of the BESIII detector consists of a helium-based
 multilayer drift chamber (MDC), a plastic scintillator time-of-flight
system (TOF), and a CsI(Tl) electromagnetic calorimeter (EMC),
which are all enclosed in a superconducting solenoidal magnet
providing a 1.0~T magnetic field. The solenoid is supported by an
octagonal flux-return yoke with resistive plate counter muon
identifier modules interleaved with steel. The acceptance of
charged particles and photons is 93\% over the $4\pi$ solid angle. The
charged-particle momentum resolution at $1~{\rm GeV}/c$ is
$0.5\%$, and the $\textrm{d}E/\textrm{d}x$ resolution is $6\%$ for the electrons
from Bhabha scattering. The EMC measures photon energies with a
resolution of $2.5\%$ ($5\%$) at $1$~GeV in the barrel (end cap)
region. The time resolution of the TOF barrel part is 68~ps, while
that of the end cap part is 110~ps. The end cap TOF
system was upgraded in 2015 with multi-gap resistive plate chamber
technology, providing a time resolution of
60~ps~\cite{etof}.

Simulated data samples produced with the {\sc geant4}-based~\cite{geant4} Monte Carlo (MC) package, which
includes the geometric description of the BESIII detector and the
detector response, are used to determine the detection efficiency
and to estimate the background contributions. The simulation models the beam
energy spread and initial state radiation (ISR) in the $e^+e^-$
annihilations using the generator {\sc kkmc}~\cite{KKMC}. For the signal we use a MC sample of the $\EE\too\omega\chi_{c0}$ process generated according to the measured angular distribution, which is introduced in section VI.
The inclusive MC samples consist of the production of open charm
processes, the ISR production of vector charmonium(-like) states,
and the continuum processes incorporated in {\sc
kkmc}~\cite{KKMC}. The known decay modes are modelled with {\sc
evtgen}~\cite{ref:evtgen} using branching fractions taken from the
Particle Data Group (PDG)~\cite{pdg}, and the remaining unknown decays
from the charmonium states are generated with {\sc
lundcharm}~\cite{ref:lundcharm}. Final state radiation (FSR) effects
from charged final state particles are incorporated $via$ the {\sc
photos} package~\cite{photos}.

\section{III. EVENT SELECTION}
For each charged track, the distance of closest approach to the interaction point (IP) is required to be within $\pm10$ cm in the beam direction and within 1 cm in the plane perpendicular to the beam direction. The polar angle ($\theta$) of the tracks must be within the fiducial volume of the MDC $(|\cos\theta|<0.93)$. Photons are reconstructed from isolated showers in the EMC, which are at least $10^\circ$ away from the nearest charged track. The photon energy is required to be at least 25 MeV in the barrel region $(|\cos\theta|<0.8)$ or 50 MeV in the end cap region $(0.86<|\cos\theta|<0.92)$. To suppress electronic noise and energy depositions unrelated to the event, the time after the collision at which the photon is recorded in the EMC is required to satisfy $0\leq t \leq 700$ ns.

Since the final states of the $\EE \too \omega \chi_{c0}$ signal are $\pi^{0}\pi^{+}\pi^{-}\pi^{+}\pi^{-}$ or $\pi^{0}\pi^{+}\pi^{-}K^{+}K^{-}$, candidate events must have four tracks with zero net charge and at least two photons. The tracks with a momentum larger than 1~GeV/$c$ are identified as $\pi/K$ from the decay of the $\chi_{c0}$, whereas lower momentum tracks are considered as pions from $\omega$ decays. Since the tracks from $\omega$ and $\chi_{c0}$ can be separated clearly according to the momentum, the mis-identification rate is negligible. To reduce the background contributions and to improve the mass resolution, a 5C-kinematic fit is performed to both constrain the total four momentum of the final state particles to the total initial four momentum of the colliding beams and to constrain the invariant mass of the two photons from the decay of the $\pi^{0}$ to its nominal mass~\cite{pdg}. If there is more than one candidate in an event, the average multiplicity for signal is 1.09, the one with the smallest $\chi^{2}_{\text{5C}}$ of the kinematic fit is selected. The two track candidates of the decay of the $\chi_{c0}$ are considered to be either a $\pi^{+}\pi^{-}$ or a $K^{+}K^{-}$ pair depending on the $\chi^2$ of the 5C-kinematic fit. If $\chi^{2}_{\text{5C}}(\pi^{+}\pi^{-}) < \chi^{2}_{\text{5C}}(K^{+}K^{-})$, the two tracks are identified as a $\pi^{+}\pi^{-}$ pair, otherwise, as a $K^{+}K^{-}$ pair. The $\chi^{2}_{\text{5C}}$ of the candidate events is required to be less than 100.

\section{IV. BORN CROSS SECTION MEASUREMENT}
\begin{figure}[htbp]
\begin{center}
\includegraphics[width=0.43\textwidth]{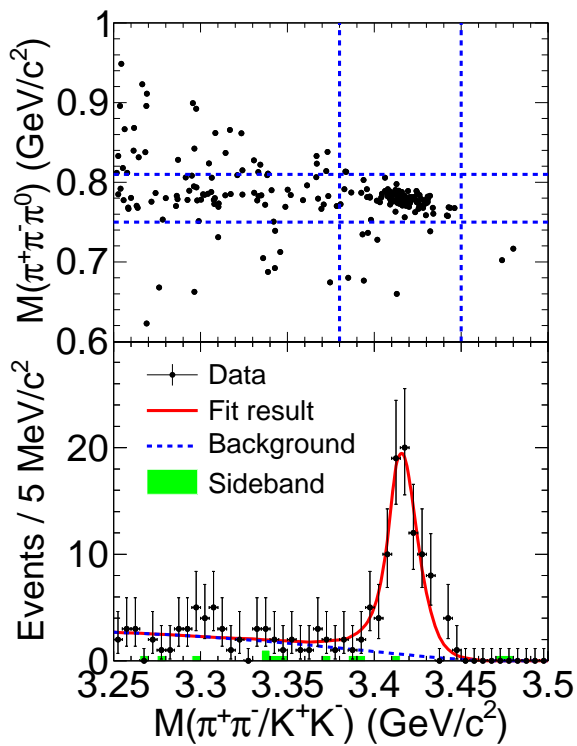}
\caption{ (top) The distribution of $M(\pi^{+}\pi^{-}\pi^{0})$ versus $M(\pi^{+}\pi^{-}/K^{+}K^{-})$ for data at $\sqrt{s}=4.219$ GeV.  The blue dashed lines denote the $\omega$ and $\chi_{c0}$ mass bands. (bottom) The invariant mass $M(\pi^{+}\pi^{-}/K^{+}K^{-})$ distribution for the data at $\sqrt{s}=4.219$ GeV. The red solid line is the fit to the data and the blue dashed line is a fit of the background. The green shaded histogram corresponds to the normalized background events from the $\omega$ sideband region.}
\label{fig:scatter}
\end{center}
\end{figure}

The correlation between the $\pi^{+}\pi^{-}\pi^{0}$ invariant mass, $M(\pi^{+}\pi^{-}\pi^{0})$, and the $\pi^{+}\pi^{-}/K^{+}K^{-}$ mass, $M(\pi^{+}\pi^{-}/K^{+}K^{-})$, is shown in the top panel in Fig.~\ref{fig:scatter} for data taken at $\sqrt{s} = 4.219$ GeV. A high density area can be observed that originates from the $\EE \too \omega\chi_{c0}$ channel. The mass range [0.75, 0.81]~GeV/$c^{2}$ in $M(\pi^{+}\pi^{-}\pi^{0})$ is defined as the $\omega$ signal region and is indicated by horizontal dashed lines. A sideband in the range [0.60, 0.72] GeV$/c^{2}$ is used to study the non-resonant background. The $\chi_{c0}$ signal region is indicated by the vertical dashed lines and is defined as [3.38, 3.45] GeV/$c^{2}$. The bottom panel of Fig.~\ref{fig:scatter} shows the distribution of $M(\pi^{+}\pi^{-}/K^{+}K^{-})$ for data in the $\omega$ signal region. The shaded (green) histogram corresponds to normalized events in the $\omega$ sideband region.

\begin{table*}[htbp]
\begin{center}
  \caption{Born cross sections $\sigma^{\rm B}$ (or upper limits at 90\% C.L. $\sigma^{\rm B}_{\text{upper}}$) for the $e^+e^-\to \omega \chi_{c0}$ reaction at the different center-of-mass energies $\sqrt{s}$, together with integrated luminosities $\mathcal{L}$, number of signal events $N^{\rm sig}$, radiative correction factor 1 + $\delta$(s), vacuum polarization factor $\frac{1}{|1-\Pi|^{2}}$, and efficiency $\epsilon$. The first uncertainties are statistical, and the second systematic.}
\label{tab:crosssection}
\begin{tabular}{c  c  c  c  c  c  c }
  \hline
  \hline
  \ \ \ \ $\sqrt{s}$ (GeV) \ \ \ \ & \ \ \ \ $\mathcal{L}$ (pb$^{-1}$) \ \ \ \ & \ \ \ \ $N^{\rm sig}$ \ \ \ \ & \ \ \ \ 1 + $\delta$(s) \ \ \ \ & \ \ \ \ $\frac{1}{|1-\Pi|^{2}}$ \ \ \ \ & \ \ \ \ $\epsilon$ $(\%)$ \ \ \ \ & \ \ \ \ $\sigma^{\rm B}(\sigma^{\rm B}_{\text{upper}})$ (pb) \ \ \ \  \\
  \hline
  4.178 & 3194.5 & $0.0^{+11.6}_{-0.0}$ & 0.63 & 1.055 & 24.71 & $0.0^{+2.2+0.5}_{-0.0-0.0}(<4.0)$  \\
  4.189 & 524.6 & $5.4\pm4.7$ & 0.64 & 1.056 & 24.59 & $6.2\pm5.4\pm1.1(<15)$  \\
  4.199 & 526.0 & $21.5\pm6.4$ & 0.66 & 1.057 & 25.68 & $22.6\pm6.7\pm2.6$  \\
  4.209 & 518.0 & $27.8\pm8.4$ & 0.68 & 1.057 & 25.70 & $28.8\pm8.7\pm4.3$  \\
  4.219 & 514.6 & $92.5\pm11.2$ & 0.71 & 1.057 & 25.52 & $93.0\pm11.3\pm8.5$  \\
  4.236 & 530.3 & $61.3\pm9.9$ & 0.80 & 1.056 & 25.92 & $52.2\pm8.4\pm4.7$  \\
  4.244 & 538.1 & $21.9\pm8.0$ & 0.86 & 1.055 & 25.51 & $17.4\pm6.4\pm2.5$  \\
  4.267 & 531.1 & $12.7\pm9.1$ & 1.44 & 1.053 & 22.10 & $7.1\pm5.1\pm1.7(<16)$  \\
  4.278 & 175.7 & $0.0^{+3.0}_{-0.0}$ & 2.68 & 1.053 & 17.11 & $0.0^{+3.5+0.9}_{-0.0-0.0}(<6.8)$  \\
  \hline
  \hline
\end{tabular}
\end{center}
\end{table*}

An unbinned maximum likelihood fit is performed to obtain the signal yields. In the fit, we use the MC-determined shape to describe the $\chi_{c0}$ signal. The background is described with a generalized ARGUS function~\cite{argus}
\begin{equation}
m\cdot(1-(\frac{m}{m_{0}})^{2})^{p}\cdot{\text{exp}(k(1-(\frac{m}{m_{0}})^{2}))}\cdot\theta(m-m_{0}) ,
\end{equation}
where $m_{0}$ is fixed to $(\sqrt{s}-0.75$ GeV), with 0.75 GeV being the lower limit of $M(\pi^{+}\pi^{-}\pi^{0})$, and $p$, $k$ are free parameters. The red solid curve in the bottom panel of Fig.~\ref{fig:scatter} shows the fit result. The data taken at the other center-of-mass energies are analyzed using the same method and the fit results are summarized in Table~\ref{tab:crosssection}.

The Born cross section is calculated with
\begin{equation}
    \sigma^{\rm B}(\EE \rightarrow \omega \chi_{c0}) = \frac{N^{\rm sig}}{\mathcal{L}(1 + \delta (s))\frac{1}{|1-\Pi|^{2}}{\mathcal{B}}{\epsilon}} ,
\end{equation}
where $N^\text{sig}$ is the number of signal events, $\mathcal{L}$ is the integrated luminosity obtained using the same method in Ref.~\cite{luminosity}, 1 + $\delta (s)$ is the radiative correction factor obtained from a Quantum Electrodynamics (QED) calculation~\cite{QED, KKMC} using the obtained preliminary cross section as input and iterating it until the results converge, $\frac{1}{|1-\Pi|^{2}}$ is the correction factor for vacuum polarization~\cite{vacuum}, $\mathcal{B}$ is the product of branching fractions $\mathcal{B}(\chi_{c0} \too \pi^{+}\pi^{-}/K^{+}K^{-}) \times \mathcal{B}(\omega \too \pi^{+}\pi^{-}\pi^{0}) \times \mathcal{B}(\pi^{0} \too \gamma\gamma)$, and $\epsilon$ is the event selection efficiency. The Born cross sections (or upper limits at 90\% C.L.) at each energy point for $\EE \too \omega\chi_{c0}$ are listed in Table~\ref{tab:crosssection}.

The systematic uncertainty of the Born cross section measurement originates mainly from the luminosity determination, the tracking efficiency, photon detection efficiency, kinematic fit, radiative correction, fit range, signal and background shapes, angular distribution, and the branching fractions for $\mathcal{B}(\chi_{c0} \too \pi^{+}\pi^{-}/K^{+}K^{-}) \times \mathcal{B}(\omega \too \pi^{+}\pi^{-}\pi^{0}) \times \mathcal{B}(\pi^{0} \too \gamma\gamma)$.

The luminosity is measured with a precision of about $1.0\%$ using the well-known Bhabha scattering process~\cite{luminosity}. The uncertainty in the tracking efficiency is obtained as $1.0\%$ per track using the process $\EE \too \pi^+\pi^-K^+K^-$~\cite{omegachic2}. The uncertainty in photon reconstruction is $1.0\%$ per photon, obtained by studying the $J/\psi \too \rho^{0}\pi^{0}$ decay~\cite{photon}.

The systematic uncertainty due to the kinematic fit is estimated by correcting the helix parameters of charged tracks according to the method described in Ref.~\cite{helix}. The difference between detection efficiencies obtained from MC samples with and without correction is taken as the uncertainty.

The line-shape of the $\EE \too \omega \chi_{c0}$ cross section will affect the radiative correction factor and the efficiency. In the nominal results, we use a bifurcated Gaussian function as the line-shape to describe the cross section. The shape is used as input and is iterated until the results converge. To estimate the uncertainty from the radiative correction, we change the line-shape to the Breit-Wigner (BW) function of the $Y(4220)$~\cite{omegachic2}. The difference between the results is taken as a systematic uncertainty.

The uncertainty from the fit range is obtained by varying the limits of the fit range by $\pm$0.01 GeV/$c^{2}$. We take the largest difference of the corresponding cross section measurement with respect to the nominal one as the systematic uncertainty.  We use the MC-determined shape convolved with a Gaussian function to fit the data as input to get the uncertainty of the signal shape. The difference in the results with respect to the nominal one is taken as the systematic uncertainty. To estimate the systematic uncertainty caused by the background shape, we vary $m_{0}$ by $\pm0.01$ GeV/$c^2$ in the ARGUS function, and take the largest difference in the results as the uncertainty.

\begin{table*}[htbp]
\begin{center}
\caption{Relative systematic uncertainties (in $\%$) from the different sources. Sources marked with an asterisk have common relative systematic uncertainties for the different center-of-mass energies. Dashes mean that the results are not applicable. }
\label{tab:sumerror}
\begin{tabular}{c c c c c c c c c c c}
  \hline
  \hline
  \ \ Source / $\sqrt{s}$ (GeV)  \ \ & \ \ 4.178 \ \ & \ \ 4.189 \ \ & \ \ 4.199 \ \ & \ \ 4.209 \ \ & \ \ 4.219 \ \ & \ \ 4.236 \ \ & \ \ 4.244 \ \ & \ \ 4.267 \ \ & \ \ 4.278 \ \ \\
  \hline
  Luminosity$^{*}$ & 1.0 & 1.0 & 1.0 & 1.0 & 1.0 & 1.0 & 1.0 & 1.0 & 1.0 \\
  Tracking efficiency$^{*}$ & 4.0 & 4.0 & 4.0 & 4.0 & 4.0 & 4.0 & 4.0 & 4.0 & 4.0 \\
  Photon detection$^{*}$ & 2.0 & 2.0 & 2.0 & 2.0 & 2.0 & 2.0 & 2.0 & 2.0 & 2.0 \\
  Kinematic fit & 0.6 & 0.8 & 0.6 & 0.6 & 0.5 & 0.6 & 0.8 & 2.4 & 3.8  \\
  Radiative correction & 14.0 & 11.2 & 4.5 & 3.6 & 0.2 & 0.9 & 1.8 & 19.8 & 45.9 \\
  Fit range & $-$ & 9.3 & 2.3 & 2.9 & 2.5 & 2.0 & 2.3 & 8.7 & $-$ \\
  Signal shape & $-$ & 1.9 & 5.1 & 9.4 & 0.2 & 1.5 & 11.4 & 1.6 & $-$ \\
  Background shape & $-$ & 1.9 & 3.7 & 6.5 & 2.8 & 1.5 & 2.3 & 6.3 & $-$ \\
  Angular distribution & 0.1 & 0.1 & 0.3 & 0.5 & 0.5 & 0.8 & 0.6 & 0.7 & 0.9 \\
  Branching fraction$^{*}$ & 6.9 & 6.9 & 6.9 & 6.9 & 6.9 & 6.9 & 6.9 & 6.9 & 6.9 \\
  Sum & 16.3 & 17.0 & 11.6 & 14.9 & 9.2 & 8.9 & 14.6 & 24.2 & 46.8   \\
  \hline
  \hline
\end{tabular}
\end{center}
\end{table*}

The measured angular distribution is used as a model to generate signal events in the MC simulations. The detection efficiency of the $\EE \too \omega \chi_{c0}$ reaction will depend upon its angular distribution. We obtained an angular distribution parameter, defined in section VI, of $\alpha=-0.30\pm0.18(\text{stat.})\pm0.05(\text{sys.})$. The systematic uncertainty of the efficiency due to uncertainties in the angular distribution is estimated by varying the $\alpha$ value by one standard deviation, the total uncertainty on $\alpha$.

The uncertainty in the product of the branching fractions $\mathcal{B}(\chi_{c0} \too \pi^{+}\pi^{-}/K^{+}K^{-}) \times \mathcal{B}(\omega \too \pi^{+}\pi^{-}\pi^{0}) \times \mathcal{B}(\pi^{0} \too \gamma\gamma)$ is taken from the uncertainties quoted by the PDG~\cite{pdg}.

Table~\ref{tab:sumerror} summarizes all the systematic uncertainties related to the cross section measurements of the $\EE \too \omega \chi_{c0}$ process for each center-of-mass energy. The overall systematic uncertainties are obtained by adding all the sources of systematic uncertainties in quadrature assuming they are uncorrelated.

\section{V. resonant parameter measurement}
\begin{figure}[htbp]
\begin{center}
\includegraphics[width=0.43\textwidth]{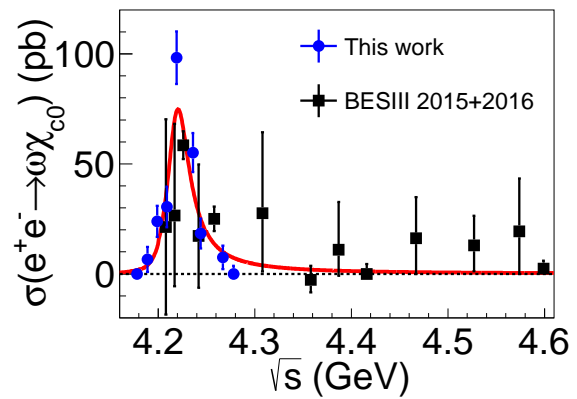}
\caption{ The  $\EE \too \omega \chi_{c0}$ cross section as a function of the center-of-mass energy. The blue points are from this work, the black square points are from Refs.~\cite{omegachic, omegachic2} and the red solid line is the fit result. }
\label{fig:crosssection}
\end{center}
\end{figure}

Figure~\ref{fig:crosssection} shows the dressed cross sections ($\sigma=\frac{\sigma^{\rm B}}{|1-\Pi|^2}$) for the $\EE \too \omega \chi_{c0}$ reaction as a function of center-of-mass energy. The black square points are taken from Refs.~\cite{omegachic, omegachic2}, and the blue circular points are from this work. We observe an enhancement in the cross section around 4.22 GeV. By assuming that the $\omega \chi_{c0}$ signals all come from a single resonance, which we label as the $Y(4220)$, with mass $M$ and width $\Gamma$, we fit the cross section data with the following formula convolved with a Gaussian function for the energy spread:
\begin{equation}
    \sigma(\sqrt{s}) = \frac{12\pi\Gamma_{ee}\mathcal{B}(\omega \chi_{c0})\Gamma}{(s-M^{2})^2+M^2\Gamma^2}\times\frac{\Phi(\sqrt{s})}{\Phi(M)},
\end{equation}
where $\Phi(\sqrt{s})$ is the two-body phase space factor and $\Gamma_{ee}$ is the electronic width. The fit to all the data in Fig.~\ref{fig:crosssection} gives $\Gamma_{ee}\mathcal{B}(\omega\chi_{c0})=(2.5\pm0.2)$ eV, $M=(4218.5\pm1.6)$~MeV/$c^{2}$, $\Gamma=(28.2\pm3.9)$ MeV, where the uncertainties are statistical only. In the fit, the cross sections' statistical uncertainties are used only. The goodness of fit is $\chi^2/\text{ndf}=29/19$, where $\text{ndf}$ is the number of degrees of freedom.

The systematic uncertainties on the resonant parameters mainly arise from uncertainties in the absolute beam energy, the parametrization of the BW function, and the cross section measurement. Because the energy spread effect has been considered in the fit, we ignore the systematic uncertainty from energy spread.

Since the uncertainty of the beam energy is about 0.8~MeV, which is obtained using the same method in Ref.~\cite{ecms}, the uncertainty of the resonant parameters caused by the beam energy is estimated by varying $\sqrt{s}$ within 0.8 MeV.

The cross section has been fitted with a BW function having the energy-dependent width $\Gamma=\Gamma^{0}\frac{\Phi(\sqrt{s})}{\Phi(M)}$ in the denominator, where $\Gamma^{0}$ is the nominal width of the resonance, to estimate the uncertainty from parametrization of the BW function. The difference between this fit result and the nominal result is taken as the uncertainty from the parametrization of the BW function.

The systematic uncertainty of the cross section measurement will affect the resonant parameters in the fit and can be divided into two parts. One part comes from the uncorrelated uncertainty among the different center-of-mass energies, and the other part is a common uncertainty. The first part has been considered by including the systematic uncertainty of the cross section in the fit. The difference between the parameters obtained in this fit to those from the nominal fit is taken as the uncertainty. We vary the cross section within the systematic uncertainty coherently for the second part and take the difference between this fit result and the nominal result as the uncertainty. We add the two parts in quadrature assuming they are uncorrelated.

Table~\ref{tab:sumfiterror} summarizes all the systematic uncertainties of the resonant parameters. The total systematic uncertainty is obtained by summing all the sources of systematic uncertainties in quadrature by assuming they are uncorrelated.
\begin{table}[htbp]
\begin{center}
\caption{Summary of systematic uncertainties on the resonant parameters. The units for $\Gamma_{ee}\mathcal{B}(\omega\chi_{c0})$, $M$, and $\Gamma$ are eV, MeV/$c^{2}$, and MeV, respectively. }
\label{tab:sumfiterror}
\begin{tabular}{c c c c}
  \hline
  \hline
  \ \ & \ \ $\Gamma_{ee}\mathcal{B}(\omega\chi_{c0})$ \ \ & \ \ $M$ \ \ & \ \ $\Gamma$  \ \ \\
  \hline
  Absolute beam energy & 0.1 & 0.9 & 0.2 \\
  Resonance parametrization & 0.1 & 3.9 & 1.1 \\
  Cross section measurement & 0.2 & 0.3 & 1.1 \\
  Sum & 0.3 & 4.0 & 1.6 \\
  \hline
  \hline
\end{tabular}
\end{center}
\end{table}

\section{VI. angular distribution measurement}
\begin{figure}[htbp]
\begin{center}
\begin{overpic}[width=0.23\textwidth]{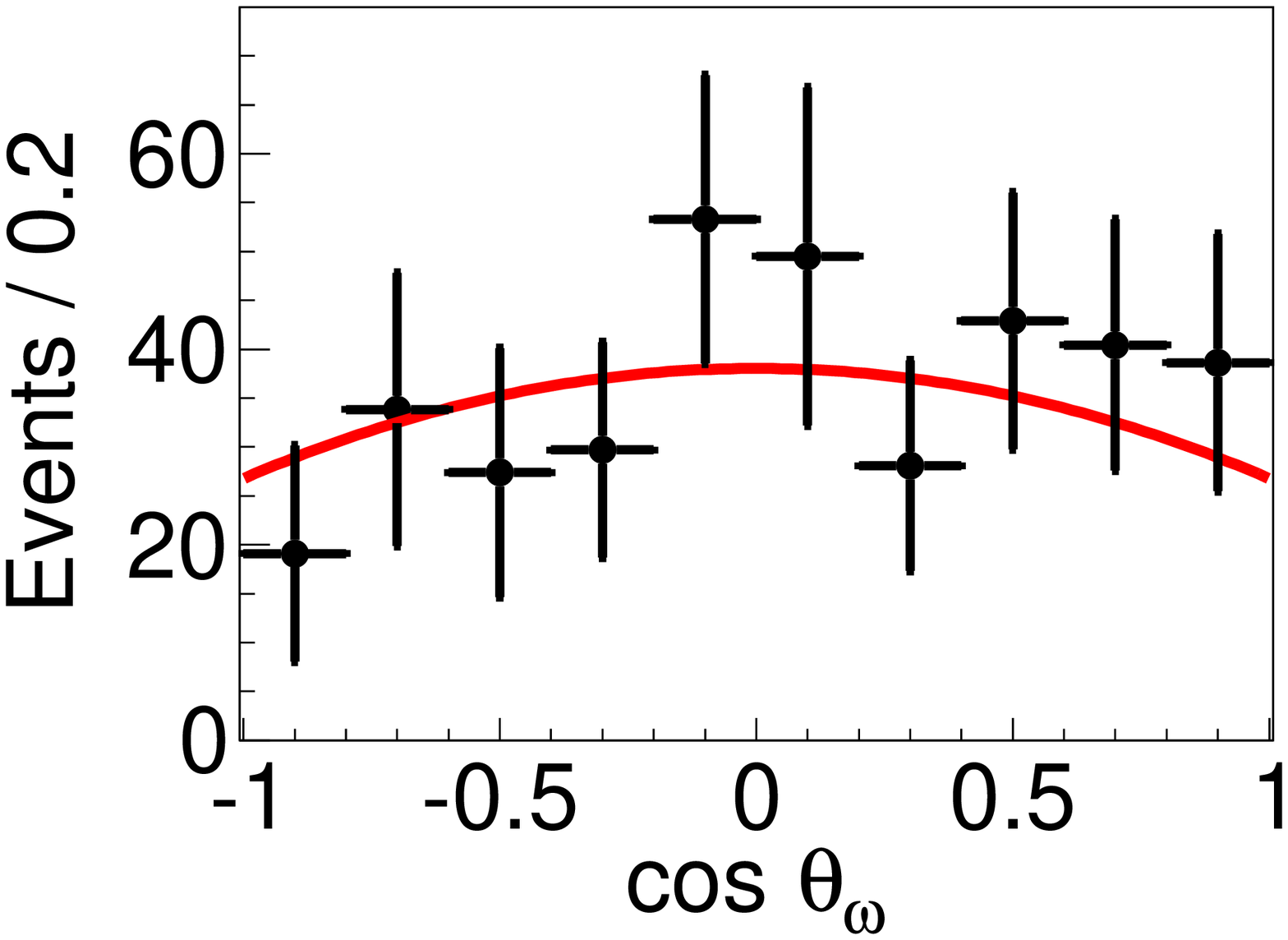}
\put(97,70){(a)}
\end{overpic}
\begin{overpic}[width=0.23\textwidth]{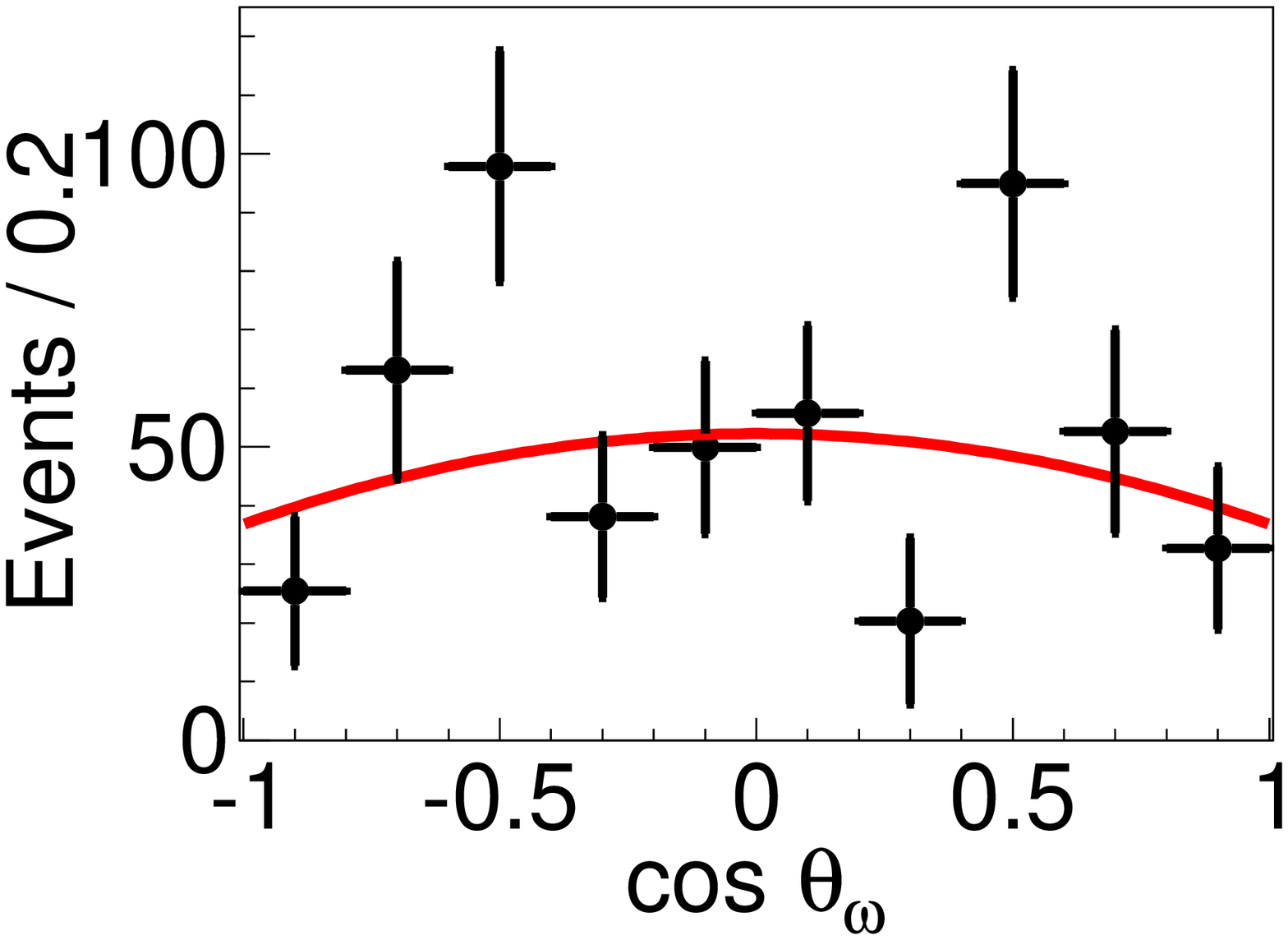}
\put(97,70){(b)}
\end{overpic}
\begin{overpic}[width=0.23\textwidth]{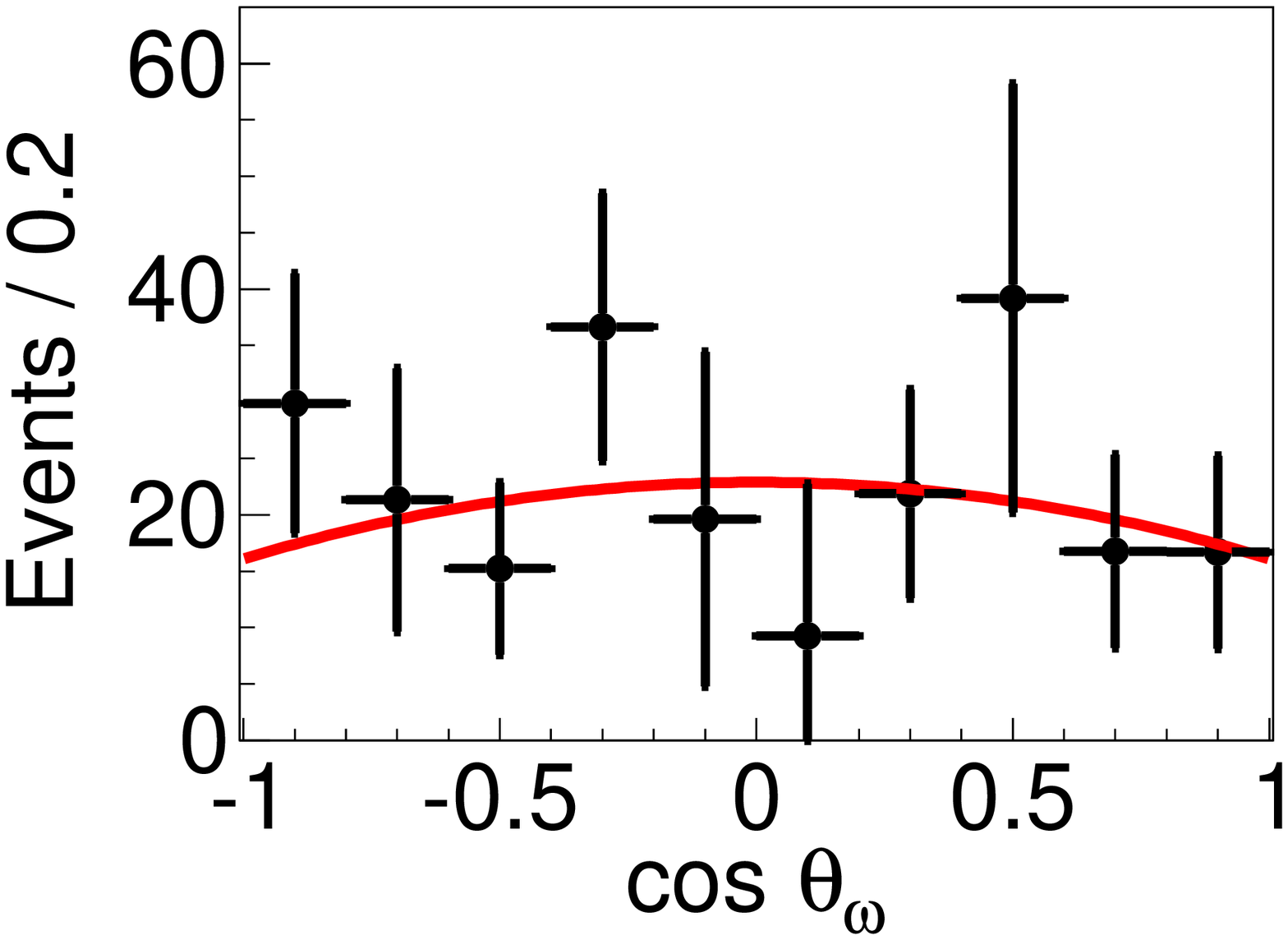}
\put(97,70){(c)}
\end{overpic}
\begin{overpic}[width=0.23\textwidth]{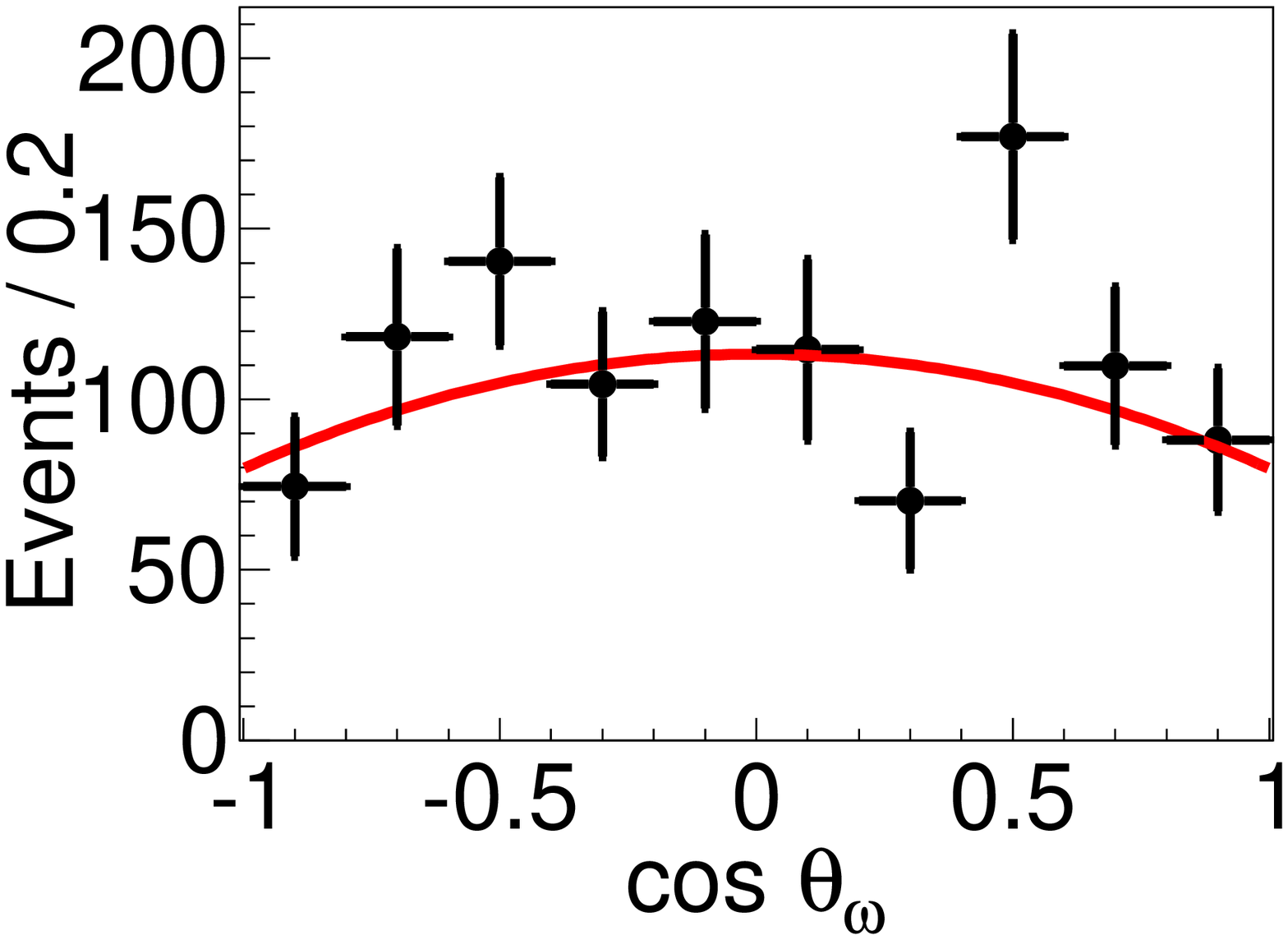}
\put(97,70){(d)}
\end{overpic}
\caption{Simultaneous fit to the angular distributions for data taken at $\sqrt{s}$ = 4.219 (a), 4.226 (b) and 4.236 (c) GeV. (d) shows the summed result of the three center-of-mass energies.}
\label{fig:angle}
\end{center}
\end{figure}

Both $S$ and $D$-wave contributions are possible in the process $Y(4220) \too \omega \chi_{c0}$. A measurement of their strengths can be helpful to extract information about the underlying dynamics of the decay process. We therefore performed an angular analysis~\cite{angular} of the relatively high-statistics data samples taken at $\sqrt{s}$ = 4.219, 4.226, and 4.236 GeV (selection of $\sqrt{s}$ = 4.226 GeV data was reported in Ref.~\cite{omegachic}). The helicity angle, $\theta_{\omega}$, defined by the scattering angle of the $\omega$ with respect to the electron beam in the $\EE$ center-of-mass frame was reconstructed for each event. Figure~\ref{fig:angle} shows the bin-by-bin efficiency-corrected events as a function of $\text{cos}\theta_{\omega}$ for the three center-of-mass energies. The signal yield in each of the 10 bins is determined with the same method as that in the cross section measurement, and the detection efficiency in each bin is determined with the signal MC sample. We performed a simultaneous fit using the function $1+\alpha$cos$^{2}\theta_{\omega}$ with a least-square method, assuming $\alpha$ is common to the three energy points. The red line in Fig.~\ref{fig:angle} shows the best fit result with $\alpha=-0.30\pm0.18\pm0.05$, where the first uncertainty is statistical and the second systematic. The goodness of the fit is $\chi^2/$ndf$=31/26$. The fit result indicates evidence for a combination of $S$ and $D-$wave contributions in the $Y(4220) \too \omega \chi_{c0}$ process, although the statistical significance of this conclusion is only $2\sigma$ compared with a pure $S-$wave contribution of $\alpha=0$.

The systematic uncertainty of $\alpha$ has been estimated by varying the fit range (0.02), the signal (0.01) and background shapes (0.03), and the radiative correction factor (0.03). The uncertainties are indicated in brackets and are determined with the same method described earlier for the cross section measurements. In addition, we estimate an additional source of systematic uncertainty by varying the number of bins. For this, we change the number of bins from 10 to 8, and repeat the process. The difference in $\alpha$ is found to be 0.01. The overall systematic uncertainty (0.05) is obtained by summing all the items of systematic uncertainties in quadrature by assuming they are uncorrelated.

\section{VII. Summary}
The process $\EE \too \omega \chi_{c0}$ has been studied using 9 data samples collected at center-of-mass energies from $\sqrt{s} =$ 4.178 to 4.278 GeV. The $\sqrt{s}$-dependence of the cross section has been measured and the results are listed in Table~\ref{tab:crosssection} and are shown in Fig.~\ref{fig:crosssection}. A clear enhancement is seen around $\sqrt{s} =$ 4.22 GeV which confirms, and statistically improves upon, an earlier observation~\cite{omegachic}. By fitting the $\EE \too \omega \chi_{c0}$ cross section with a single resonance, the mass and width for the structure are determined to be $M=(4218.5\pm1.6(\text{stat.})\pm4.0(\text{sys.}))$ MeV/$c^2$ and $\Gamma=(28.2\pm3.9(\text{stat.})\pm1.6(\text{sys.}))$ MeV. The obtained resonance parameters are not compatible with the vector charmonium state $\psi(4160)$, ruling out its possible contribution to the structure~\cite{the4}. Moreover, we studied the angular distribution of the process $Y(4220) \too \omega \chi_{c0}$. We measured $\alpha=-0.30\pm0.18\pm0.05$, which indicates a combination of $S$ and $D-$wave contributions in the decay.

\begin{figure}[htbp]
\begin{center}
\includegraphics[width=0.47\textwidth]{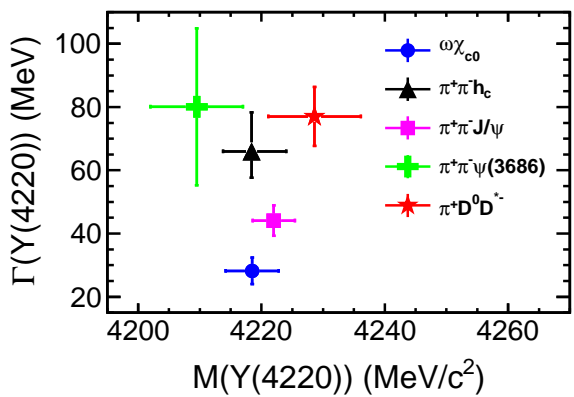}
\caption{ Mass and width of the $Y(4220)$ obtained from the processes $\EE \too \omega\chi_{c0}$, $\pi^+\pi^-h_c$, $\pi^+\pi^-J/\psi$, $\pi^+\pi^-\psi(3686)$ and $\pi^{+}D^{0}D^{*-}+c.c.$}
\label{fig:y4220}
\end{center}
\end{figure}

Figure~\ref{fig:y4220} shows the measured mass and width of the $Y(4220)$ from the different processes. The masses are consistent with each other, while the widths are not. The widths from the processes $\EE \too \pi^+\pi^-h_c$, $\pi^+\pi^-\psi(3686)$, and $\pi^{+}D^{0}D^{*-}+c.c.$ are larger than those from the processes $\EE \too \omega\chi_{c0}$ and $\pi^+\pi^-J/\psi$. From these inconsistencies in the width, we cannot draw a conclusion on whether the structure observed in these processes is the same state or whether the inconsistencies are caused by the BW parameterization. Further experimental studies with higher statistics are needed to draw a more reliable conclusion on the nature of this structure.

\section{ACKNOWLEDGMENTS}
The BESIII collaboration thanks the staff of BEPCII and the IHEP computing center for their strong support. This work is supported in part by National Key Basic Research Program of China under Contract No. 2015CB856700; National Natural Science Foundation of China (NSFC) under Contracts Nos. 11625523, 11635010, 11735014; National Natural Science Foundation of China (NSFC) under Contract No. 11835012; National Natural Science Foundation of China (NSFC) under Contract No. 11847028; National Natural Science Foundation of China (NSFC) under Contract No. 11575198; the Chinese Academy of Sciences (CAS) Large-Scale Scientific Facility Program; Joint Large-Scale Scientific Facility Funds of the NSFC and CAS under Contracts Nos. U1532257, U1532258, U1732263, U1832207; CAS Key Research Program of Frontier Sciences under Contracts Nos. QYZDJ-SSW-SLH003, QYZDJ-SSW-SLH040; 100 Talents Program of CAS; INPAC and Shanghai Key Laboratory for Particle Physics and Cosmology; Foundation of Henan Educational Committee (No. 19A140015); Nanhu Scholars Program for Young Scholars of Xinyang Normal University; German Research Foundation DFG under Contract No. Collaborative Research Center CRC 1044; Istituto Nazionale di Fisica Nucleare, Italy; Koninklijke Nederlandse Akademie van Wetenschappen (KNAW) under Contract No. 530-4CDP03; Ministry of Development of Turkey under Contract No. DPT2006K-120470; National Science and Technology fund; The Swedish Research Council; U. S. Department of Energy under Contracts Nos. DE-FG02-05ER41374, DE-SC-0010118, DE-SC-0012069; University of Groningen (RuG) and the Helmholtzzentrum fuer Schwerionenforschung GmbH (GSI), Darmstadt.


\begin{thebibliography}{**}

\bibitem{Y4260-babar} B.~Aubert {\it et al.} [BaBar Collaboration], Phys.\ Rev.\ Lett.\  {\bf 95}, 142001 (2005).

\bibitem{Y4260-cleo} Q.~He {\it et al.} [CLEO Collaboration], Phys.\ Rev.\ D {\bf 74}, 091104 (2006).

\bibitem{Y4260-belle} C.~Z.~Yuan {\it et al.} [Belle Collaboration], Phys.\ Rev.\ Lett.\  {\bf 99}, 182004 (2007).

\bibitem{Y4360-babar} B.~Aubert {\it et al.} [BaBar Collaboration], Phys.\ Rev.\ Lett.\  {\bf 98}, 212001 (2007).

\bibitem{Y4360-belle} X.~L.~Wang {\it et al.} [Belle Collaboration], Phys.\ Rev.\ Lett.\  {\bf 99}, 142002 (2007).

\bibitem{Y-theory1} N.~Brambilla {\it et al.}, Eur.\ Phys.\ J.\ C {\bf 71}, 1534 (2011).

\bibitem{Y-theory2} R.~A.~Briceno {\it et al.}, Chin.\ Phys.\ C {\bf 40}, 042001 (2016).

\bibitem{Y-theory3} H.~X.~Chen, W.~Chen, X.~Liu and S.~L.~Zhu, Phys.\ Rept.\  {\bf 639}, 1 (2016).

\bibitem{omegachic} M.~Ablikim {\it et al.} [BESIII Collaboration], Phys.\ Rev.\ Lett.\  {\bf 114}, 092003 (2015).

\bibitem{omegachic2} M.~Ablikim {\it et al.} [BESIII Collaboration], Phys.\ Rev.\ D {\bf 93}, 011102 (2016).

\bibitem{the1} L.~Y.~Dai, M.~Shi, G.~Y.~Tang and H.~Q.~Zheng, Phys.\ Rev.\ D {\bf 92}, 014020 (2015).

\bibitem{the2} L.~Ma, X.~H.~Liu, X.~Liu and S.~L.~Zhu, Phys.\ Rev.\ D {\bf 91}, 034032 (2015).

\bibitem{the3} D.~Y.~Chen, X.~Liu and T.~Matsuki, Phys.\ Rev.\ D {\bf 91}, 094023 (2015).

\bibitem{the4} X.~Li and M.~B.~Voloshin, Phys.\ Rev.\ D {\bf 91}, 034004 (2015).

\bibitem{the5} R.~Faccini, G.~Filaci, A.~L.~Guerrieri, A.~Pilloni and A.~D.~Polosa, Phys.\ Rev.\ D {\bf 91}, 117501 (2015).

\bibitem{the6} M.~Cleven and Q.~Zhao, Phys.\ Lett.\ B {\bf 768}, 52 (2017).

\bibitem{the7} Z.~G.~Wang, Chin.\ Phys.\ C {\bf 41}, 083103 (2017).

\bibitem{pipijpsi-bes} M.~Ablikim {\it et al.} [BESIII Collaboration], Phys.\ Rev.\ Lett.\ {\bf 118}, 092001 (2017).

\bibitem{pipipsip-bes} M.~Ablikim {\it et al.} [BESIII Collaboration], Phys.\ Rev.\ D {\bf 96}, 032004 (2017).

\bibitem{pipihc-bes} M.~Ablikim {\it et al.} [BESIII Collaboration], Phys.\ Rev.\ Lett.\ {\bf 118}, 092002 (2017).

\bibitem{piDDstar} M.~Ablikim {\it et al.} [BESIII Collaboration], Phys.\ Rev.\ Lett.\ {\bf 122}, 102002 (2019).

\bibitem{besiii} M.~Ablikim {\it et al.} [BESIII Collaboration], Nucl.\ Instrum.\ Meth.\ A {\bf 614}, 345 (2010).

\bibitem{bepcii} C.~H.~Yu {\it et al.}, Proceedings of IPAC2016, Busan, Korea, 2016, doi:10.18429/JACoW-IPAC2016-TUYA01.

\bibitem{etof} X.~Li {\it et al.}, Radiat. Detect. Technol. Methods {\bf 1}, 13 (2017); Y.~X.~Guo {\it et al.}, Radiat. Detect. Technol. Methods {\bf 1}, 15 (2017).

\bibitem{geant4} S.~Agostinelli {\it et al.} [GEANT4 Collaboration], Nucl.\ Instrum.\ Meth.\ A {\bf 506}, 250 (2003).

\bibitem{KKMC} S.~Jadach, B.~F.~L.~Ward and Z.~Was, Phys.\ Rev.\ D {\bf 63}, 113009 (2001); Comput.\ Phys.\ Commun.\  {\bf 130}, 260 (2000).

\bibitem{ref:evtgen} D.~J.~Lange, Nucl.\ Instrum.\ Meth.\ A {\bf 462}, 152 (2001); R.~G.~Ping, Chin. Phys. C {\bf 32}, 599 (2008).

\bibitem{pdg} M.~Tanabashi {\it et al.} [Particle Data Group], Phys.\ Rev.\ D {\bf 98}, 030001 (2018).

\bibitem{ref:lundcharm} J.~C.~Chen, G.~S.~Huang, X.~R.~Qi, D.~H.~Zhang and Y.~S.~Zhu, Phys.\ Rev.\ D {\bf 62}, 034003 (2000); R.~L.~Yang, R.~G.~Ping and H.~Chen, Chin.\ Phys.\ Lett.\  {\bf 31}, 061301 (2014).

\bibitem{photos} E.~Richter-Was, Phys.\ Lett.\ B {\bf 303}, 163 (1993).

\bibitem{argus} H.~Albrecht {\it et al.} [ARGUS Collaboration], Phys.\ Lett.\ B {\bf 241}, 278 (1990).

\bibitem{luminosity} M.~Ablikim {\it et al.} [BESIII Collaboration], Chin.\ Phys.\ C {\bf 39}, 093001 (2015).

\bibitem{QED} E.~A.~Kuraev and V.~S.~Fadin, Sov.\ J.\ Nucl.\ Phys.\  {\bf 41}, 466 (1985).

\bibitem{vacuum} S.~Actis {\it et al.} [Working Group on Radiative Corrections and Monte Carlo Generators for Low Energies Collaboration], Eur.\ Phys.\ J.\ C {\bf 66}, 585 (2010).

\bibitem{photon} M.~Ablikim {\it et al.} [BESIII Collaboration], Phys.\ Rev.\ Lett.\  {\bf 118}, 221802 (2017).

\bibitem{helix} M.~Ablikim {\it et al.} [BESIII Collaboration], Phys.\ Rev.\ D {\bf 87}, 012002 (2013).

\bibitem{ecms} M.~Ablikim {\it et al.} [BESIII Collaboration], Chin.\ Phys.\ C {\bf 40}, 063001 (2016).

\bibitem{angular} M.~Ablikim {\it et al.} [BESIII Collaboration], Phys.\ Rev.\ D {\bf 95}, 052003 (2017).

\end{thebibliography}
\end{document}